\documentclass[aps,prl,twocolumn,amsmath,amssymb,floatfix]{revtex4}
\usepackage{epsfig,psfrag,subfigure}
\usepackage{graphicx,psfrag,subfigure}
\usepackage{color}
\newcommand\beq{\begin{equation}}
\newcommand\eeq{\end{equation}}
\newcommand\bea{\begin{eqnarray}}
\newcommand\eea{\end{eqnarray}}
\newcommand\non{\nonumber}

\newcommand\de{\delta}

\newcommand\ig{\includegraphics}

\begin{document}
\title{Correlators and fractional statistics in the quantum Hall bulk}
\author{Smitha Vishveshwara$^1$}
\author{Michael Stone$^1$}
\author{Diptiman Sen$^2$}
\affiliation{$^1$ Department of Physics, University of Illinois at
Urbana-Champaign, 1110 W. Green St, Urbana, IL 61801, USA \\
$^2$ Centre for High Energy Physics, Indian Institute of Science,
Bangalore 560012, India}

\date{\today}

\begin{abstract}
We derive single-particle and two-particle correlators of anyons in the
presence of a magnetic field in the lowest Landau level. We show that the
two-particle correlator exhibits signatures of fractional statistics which can
distinguish anyons from their fermionic and bosonic counterparts. These
signatures include the zeroes of the two-particle correlator and its exclusion
behavior. We find that the single-particle correlator in finite geometries
carries valuable information relevant to experiments in which quasiparticles on
the edge of a quantum Hall system tunnel through its bulk.
\end{abstract}
\maketitle


Given the increasing interest in topological quantum computation and the rapid
experimental progress in quantum Hall physics, the study of
``anyons''---quasiparticles that obey fractional statistics interpolating
between the statistics of fermions and bosons---has gained attention not only
as a fascinating academic topic but also as one of applied value. In certain
systems\cite{leinaas,Wilczek} the detection of anyons would
establish the existence of topological order~\cite{Wen_top}. The
quantum Hall (QH) effect, where  electrons are trapped in two dimensions in the
presence of a magnetic field, provides   the  paradigm for anyon-hosting  
many-body systems~\cite{footnote}. A variety of theoretical
proposals~\cite{Kivelson,prop1,prop2,Chamon} and experimental
attempts~\cite{frac_exp} have pursued the detection of fractional statistics 
quasiparticles. While much of the understanding of anyons in these geometries
has stemmed from the edge-state description~\cite{Wen_edge} of QH
quasiparticles, an involved investigation of anyon bulk
correlations~\cite{Kivelson} has received less attention, as has the problem of 
mapping the bulk correlations  to the QH edge. In that  anyons are
intrinsically two-dimensional, a study of their bulk properties is much called
for. In this Letter  we formulate and analyze the anyon correlators in
the presence of a magnetic field in the lowest Landau level (LLL). We show that
these correlators contain valuable information on
statistics. In the presence of boundaries we demonstrate that these  bulk
correlations become manifest in edge-state properties.

\begin{figure}[t]
\ig[width=3in]{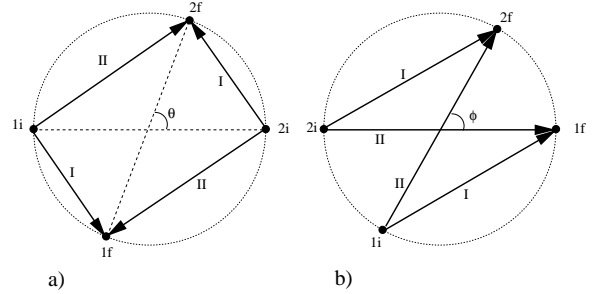}
\caption[]{Two representative configurations for anyons starting at points
$\vec{r}_{1i}$ and $\vec{r}_{2i}$ to end at points $\vec{r}_{1f}$ and
$\vec{r}_{2f}$. As the particles are indistinguishable, it is not possible to
determine which of two possible paths $I$ and $II$ each particle took.}
\vspace{-0.5cm} \label{fig:paths}
\end{figure}

The objects of our attention are the single-particle kernel $K_1(\vec{r}_f,
\vec{r}_i)$, which is the amplitude for a quasiparticle to propagate from an
initial point $\vec{r}_i$ to a final point $\vec{r}_f$, and the two-particle
kernel $K_2 (\vec{r}_{1f}, \vec{r}_{2f}, \vec{r}_{1i}, \vec{r}_{2i})$, which is
the amplitude for two quasiparticles starting at points $\vec{r}_{1i}$ and
$\vec{r}_{2i}$ to end at points $\vec{r}_{1f}$ and $\vec{r}_{2f}$ (see Fig.~\ref{fig:paths}). As elucidated in what follows, the single-particle kernel
$K_1$ lies at the heart of observable single-particle quantities such as
two-point quasiparticle correlations along a QH edge and inter-edge tunneling
matrix elements. As for $K_2$, historically, two-particle kernels have played
an ubiquitous role in a wide range of settings from particle scattering in
nuclear physics to quantum optics and astrophysics phenomena whose
study  was instigated by the
studies of Hanbury Brown and Twiss \cite{Gordon,HBT}. It is well known that the amplitude for two incoming fermions in
vacuum to scatter at an angle of $\pi/2$ is zero in the absence of a magnetic
field \cite{Gordon,Shankar} and that fermions tend to   ``anti-bunch'' while bosons
tend to ''bunch.''  Our  analysis of the kernel $K_2$ will show that the situation
is dramatically altered by the presence of a magnetic field and that
statistical effects in anyons are distinctly different from those in fermions
and bosons.

Our starting point is a two-dimensional system of two anyons in a perpendicular
magnetic field, whose common wavefunction by definition picks up a phase of
$e^{i\pi\nu}$ ($e^{-i\pi\nu}$) upon anticlockwise (clockwise) exchange of
particles. The real parameter $\nu$ lies in the range $-1 < \nu \le 1$.
The limiting cases of $\nu=0$ and $\nu=1$ correspond to bosons and fermions,
respectively. Such LLL anyon models provide
an effective description of quasihole excitations associated with the addition
of vortices to the QH bulk~\cite{Halperin,Laughlin2,Myrheim}. In particular,
for Laughlin states~\cite{Laughlin2}
quasiholes have fractional charge $q=e/m$ and anyon phase $\nu=1/m$, where $m$
is an odd integer~\cite{Laughlin1,Arovas,Wen_top}. While it is known that
interactions can alter anyon correlations in vacuum in the absence of a
magnetic field~\cite{Mottexact}, as a simple and realistic case, we treat the
anyons as non-interacting. Coulomb interactions do exist between QH quasiholes
but they are expected to be screened by the background charge and can be
treated perturbatively~\cite{Kivelson}. Thus, our treatment of the two-anyon
model ought to be applicable to QH excitations on lengthscales much larger than
the magnetic length.

The Hamiltonian for two anyons  in a perpendicular magnetic field
$\vec{B}=B\hat{z}$, in terms of center of mass and relative variables, has the
decoupled form 
\bea
H &=& \frac{1}{4\mu}\left(P_x + \frac{qB}{c} Y\right)^2 +
\frac{1}{4\mu} \left(P_y -
\frac{qB}{c} X\right)^2 \non \\
& & + \frac{1}{\mu} \left(p_x + \frac{qB}{4c} y\right)^2 + 
\frac{1}{\mu} \left(p_y -
\frac{qB}{4c} x\right)^2. \label{ham} 
\eea 
We make the particles into anyons by   requiring that when the two particles 
are  exchanged in a clockwise fashion their wavefunction
gains 
a phase factor $e^{i \pi\nu}$. Note that in an alternate
formalism, the phase factor can be gauged out of the wavefunction by treating
the anyons  as bosons carrying flux lines which can be incorporated into the
Hamiltonian \cite{Myrheim}.
Here the anyons   are assumed to each have
mass $\mu$ (which is immaterial when states are projected to the LLL) and
charge $q$. The symmetric gauge is assumed for the vector potential
$\vec{A}=B(-y\hat{x}+x\hat{y})/2$. The center of mass co-ordinate and momentum
are given by $\vec{R}=(\vec{r}_1+\vec{r}_2)/2$ and
$\vec{P}=\vec{p}_1+\vec{p}_2$ while the relative co-ordinate and momentum are
given by $\vec{r}=\vec{r}_1-\vec{r}_2$ and $\vec{p}=(\vec{p}_1-\vec{p}_2)/2$,
respectively. The Hamiltonian can also be employed to study the properties of a
single particle by restricting it to the $(\vec{R},\vec{P})$ sector where the
variables now describe the co-ordinate and momentum of the single particle.

The eigenstates of Eq. (\ref{ham}) are products of eigenstates for the center
of mass and relative coordinate systems. In the LLL, the center of mass
eigenstates are given by 
\beq \psi_n ({\vec R}) = \frac{1}{l\sqrt{m\pi n!}}
\left( \frac{Z}{\sqrt{m}} \right)^n ~\exp \left[ -\frac{|Z|^2}{2m}\right],
\label{eigcom}
\eeq 
where $n=0,1,2,\ldots$. The complex parameter $Z=(X+iY)/l$ represent the
components of $\vec{R}$, rescaled by the magnetic length $l=\sqrt{\hbar c/eB}$.
The relative-coordinate  eigenstates are given by 
\beq 
\psi_p ({\vec r})= \frac{(4\pi
m)^{-1/2}}{\sqrt{\Gamma (2p + \nu + 1)}l}  \left(\frac{z}{2\sqrt{m}}
\right)^{2p+\nu}\exp ~\left[ -\frac{|z|^2}{8m} \right], 
\label{eigrel} 
\eeq 
where
$p=0,1,2,\ldots$, and $z=(x+iy)/l$ represents rescaled components of $\vec{r}$.
These relative-coordinate  eigenstates respect the anyon property. 

We are now equipped to evaluate the single- and two-particle kernels, defined
in imaginary time $\tau$, by
\bea
 K_1 (\vec{R}_f; \vec{R}_i) &= \sum_n \psi_n (\vec{R}_f) \psi_n^*
(\vec{R}_i)~e^{-E_n \tau /\hbar}, \non \\
K_2 (\vec{r}_{1f}, \vec{r}_{2f}; \vec{r}_{1i}, \vec{r}_{2i}) &= \sum_n 
\psi_n (\vec{R}_f) \psi_n^* (\vec{R}_i)~e^{-E_n \tau /\hbar} \non \\
&\times\sum_p ~\psi_p (\vec{r}_f) ~\psi_p^* (\vec{r}_i)~e^{-E_p \tau /\hbar}.
\label{kernels} 
\eea 
In the LLL, all energies $E_n$ are degenerate, and so we  set them  to
zero. Thus  the kernels  have no  explicit time dependence. In terms of
Eqs. (\ref{eigcom},\ref{eigrel}), the single-particle kernel takes the explicit
form 
\bea 
K_1 (z_f; z_i) &=& \frac{1}{2\pi} ~\exp \left[
-\frac{1}{4} ~(|z_f|^2 + |z_i|^2) ~+~ \frac{1}{2} z_f z_i^* \right], \non \\
\label{k1} \eea and the two-particle kernel takes the form \bea & & K_2 ({\vec
r}_{1f}, {\vec r}_{2f}; {\vec r}_{1i}, {\vec r}_{2i};
\tau) \non \\
& & = \frac{1}{(2\pi m)^2} ~\exp ~\left[-\frac{1}{4m} (|z_{1f}|^2 + |z_{2f}|^2 +
|z_{1i}|^2 + |z_{2i}|^2) \right.\non \\
& & ~~~~~~~~~~~~~~~~~~~~+ \left. \frac{1}{4m} (z_{1f} + z_{2f})(z_{1i}^* +
z_{2i}^*)\right]
\non \\
& & \times ~\sum_{p=0}^\infty ~\frac{[(z_{1f} - z_{2f})(z_{1i}^* -
z_{2i}^*)/4m]^{2p+1/m}}{\Gamma (2p + 1/m +1)}. \label{k2} \eea It must be
remarked that, through different reasoning, a similar form for $K_2$ was
presented by R. Laughlin in Ref. \cite{Laughlin2}.

In the limiting case of fermions/bosons, it can be shown that the two-particle kernel can
be separated into products of individual paths, i.e., $K_2 ({\vec r}_{1f}, {\vec r}_{2f};
{\vec r}_{1i}, {\vec r}_{2i}) = K_1 ({\vec r}_{1f}; {\vec r}_{1i}) K_1 ({\vec r}_{2f},
{\vec r}_{2i})\mp K_1 ({\vec r}_{2f}; {\vec r}_{1i}) K_1 ({\vec r}_{1f}, {\vec r}_{2i})$.
From this property, several consequences follow. In particular, in the case of
Fig.~\ref{fig:paths}a, $K_2$ can only vanish if the magnitudes of the kernels along paths
of type $I$ and $II$ equal each other. This condition implies that $(z_{1f} - z_{2f})
(z_{1i}^* - z_{2i}^*)$ is imaginary, or, in other words, that $\vec{r}_{1f}-\vec{r}_{2f}$
is perpendicular to $\vec{r}_{1f}-\vec{r}_{2f}$ ($\theta=\pi/2$). Furthermore, for the two
separable parts to cancel one another, their phase difference is required to be $0/\pi$ for
fermions/bosons. This second condition translates to the requirement that the quantity
$eB\hat{z}\cdot ({\vec r}_{1i} - {\vec r}_{2i}) \times ({\vec r}_{1f} - {\vec
r}_{2f})/(hc)$ be an even/odd integer for fermions/bosons. A geometric interpretation of
these arguments is that $K_2$ vanishes when the phase difference between paths of type $I$
and $II$ is $\pi$ and that the phase is given by the sum of the Aharonov-Bohm phase picked
up by the loop in Fig.~\ref{fig:paths}. a and the phase $\pi/0$ due to anti-clockwise
exchange of the two fermions/bosons. Upon setting $\vec{B}=0$, one retrieves the result
that in the absence of a magnetic field, the two-particle kernel vanishes at an angle
$\theta=\pi/2$ for fermions.

\begin{figure}[t]
\ig[width=1.35in]{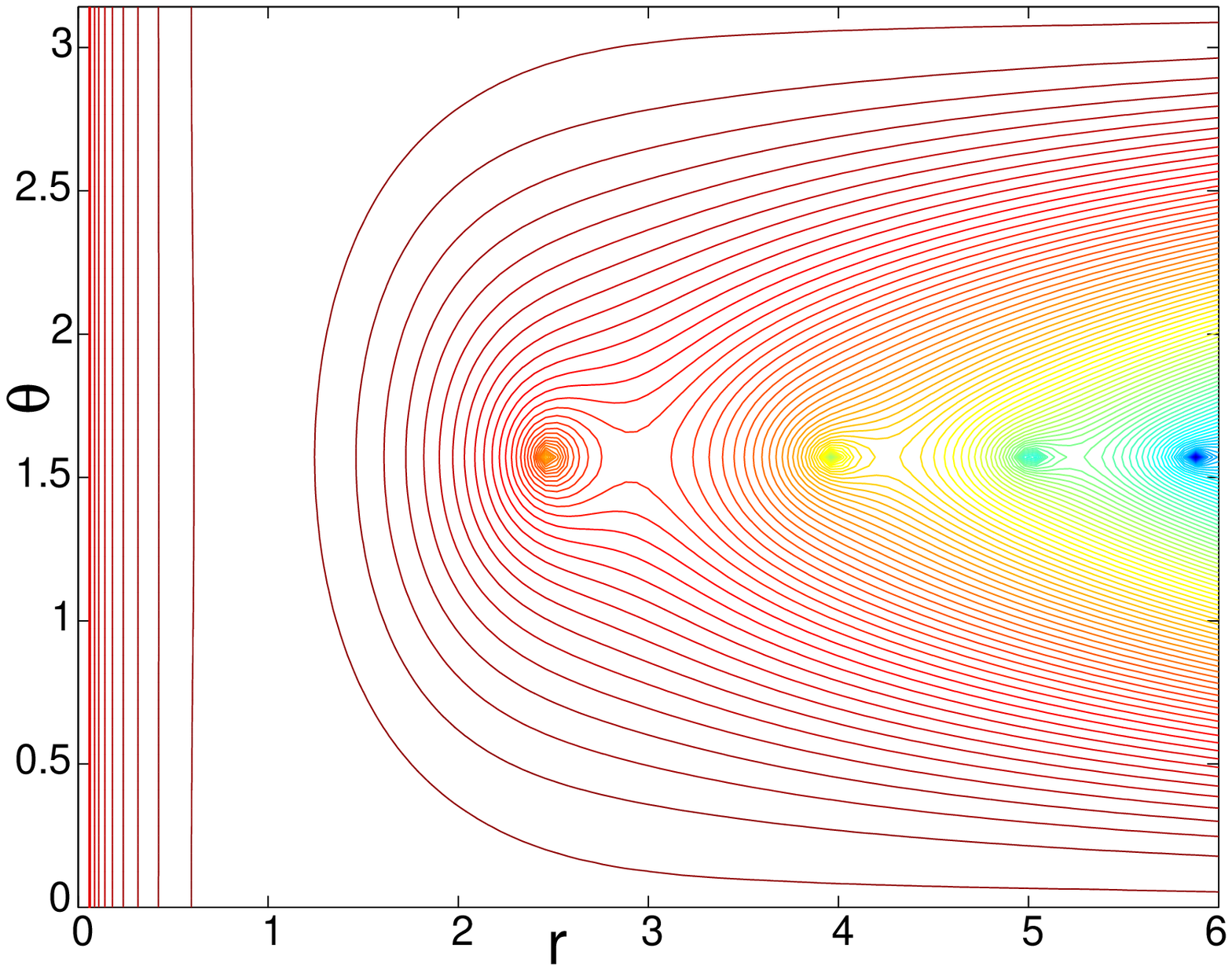}\hspace{.6cm}\ig[width=1.4in]{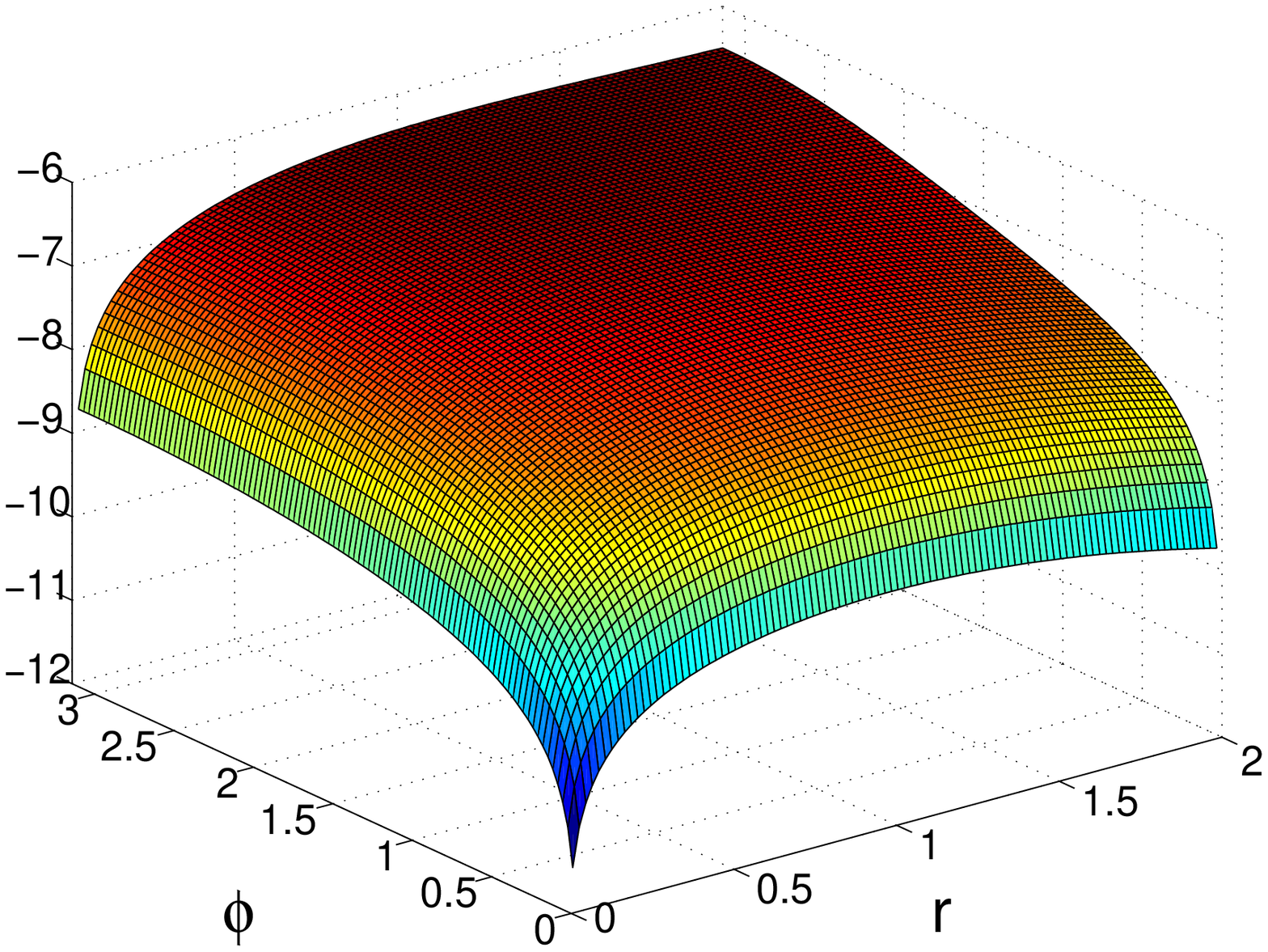}
\vspace{0.5cm}
\leftline{~~~~~~~~~~~~~~~~~~~~(a)~~~~~~~~~~~~~~~~~~~~~~~~~~~~~~~(b)}
\vspace{-1cm} \caption[]{Plots of the two-particle kernel corresponding to the
configurations in Fig. \ref{fig:paths} in which the particles lie on a circle.
The configurations correspond to (a) $z_{1i}=-z_{2i}=r$,$z_{1f}=-z_{2f}=
re^{i\theta}$, which shows zeros of the kernel at specific radii and angle
$\theta=\pi/2$ and (b) $z_{1i}=-z_{1f}=r$,$z_{2i}=-z_{2f}=re^{i\phi}$, which
shows power-law dependences of the kernel as a function of $r$ and $\phi$ for
small arguments.} \vspace{-0.5cm} \label{fig:kernel}
\end{figure}

For the case of anyons, neither the two-particle wavefunction nor the
two-particle kernel is of a separable form. However, a detailed analysis
\cite{inprep} of the two-particle kernel shows surprisingly that the geometric
arguments presented above still hold. Thus, in the configuration of Fig.
\ref{fig:paths}. a, the two-particle kernel vanishes for the same conditions
stated for fermions and bosons except that the statistical phase picked up by
the anyons for a closed loop along the $\vec{r}_{1i}\rightarrow\vec{r}_{1f}$
direction is $\pm\pi/m$ for a clockwise/anticlockwise loop. Hence, as shown in
Fig.~\ref{fig:kernel}a, the kernel vanishes along the direction
$\theta=\pi/2$ for a discrete set of radii satisfying the constraint
$r^2/m=(n-1/2+1/(2m))\pi$.

The two-particle kernel clearly exhibits features that reflect the exclusion
statistics of anyons \cite{Haldane}. As a specific instance, for the case shown
in Fig.~\ref{fig:paths}b, we find that as $\phi\rightarrow 0$, the kernel
exhibits the power-law dependence $K_2 \sim |\phi|^{2/m}$. Physically, the
amplitude for two-incoming anyons to start at nearby points and to have a small
scattering angle vanishes as the angle becomes smaller. As another instance,
the probability that two anyons are a distance $'r'$ apart is related to the
two-particle kernel whose arguments are $\vec{r}_{1i} = \vec{r}_{1f} = 0$ and
$\vec{r}_{2i} = \vec{r}_{2f} = \vec{r}$. For this case, in the limit of small
separation, the kernel has the limiting form $K_2 \rightarrow r^{2/m}$ as $r
\rightarrow 0$. For the limit $m=1$, we reproduce the result that the
probability that a fermion is at a given distance $'r'$ away from another
fermion is proportional to $r^2$. On the other hand, in the limit $m
\rightarrow \infty$, for small enough separation, one particle does not
experience the existence of another, which is indeed the situation for
condensed bosons. For any intermediate value of $m$, the power-law behavior
shows that the presence of one particle excludes that of another (thus
rendering Laughlin quasiparticles to be fermion-like), and that this
anti-bunching property becomes more pronounced for smaller values of $m$.

\begin{figure}[t]
\ig[width=3.25in]{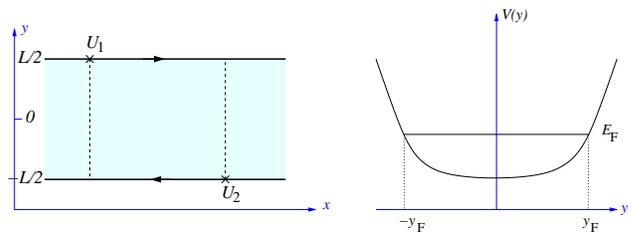}
\caption[]{Quantum Hall state in a strip geometry of length $L_x$ and width
$L_y$ in the presence of a confining potential $V(y)$. States are filled to a
Fermi energy $E_f$. Tunneling across the strip can take place via impurities
denoted by $U_i$.}
\label{fig:Hallbar}
\end{figure}

Having analyzed bulk features of the kernels, we turn to finite size
geometries that are of relevance to the physical setting of the Hall bar. By
studying the properties of the single-particle kernel in a geometry such as the
one shown in Fig.~\ref{fig:Hallbar}, we provide a simple picture for deriving
correlations along the edge and justifying assumptions made for single-particle
tunneling events in previous treatments. The system is assumed to be confined
along the $y$ direction via a potential $V(y)$. The Landau gauge $\vec{A}=-
By\hat{x}$ proves to be convenient for such a case. The corresponding
single-particle eigenstates are of the form $ \psi_{k,n} (x,y) ~=~ e^{ikx}
~f_{k,n} (y)$, where the function $f_{k,n}$ depends on the confining
potential and the momentum $k=2\pi p/L_x$ along the $x$ direction, where $p$ is
an integer and $L_x$ is the length of the strip.

We first consider the simple illustrative example of no external potential
$V(y)$ except for hard boundaries confining the strip to a width $L_y$ centered
at $y=0$. In the LLL ($n=0$), the eigenfunctions $f_{0,k}$ are Gaussian packets
proportional to $\exp [- ~\frac{eB}{2\hbar c} ~(y + \frac{\hbar ck}{eB})^2]$,
where $k$ ranges from $-eBL_y/(2\hbar c)$ to $eBL_y/(2\hbar c)$. The
single-particle kernel can be evaluated using the relationship Eq.
(\ref{kernels}). We describe the salient features of the kernel (details are
presented in Ref. \cite{inprep}). For large width $L_y \gg l$, the kernel
reproduces the behavior $K_1 \sim \exp [-(z_i-z_f)^2/4]$ when the points are
well within the bulk or on opposite edges $y_i=-y_f=L_y/2$ . In the large width
limit, if the two points however lie along the same edge, i.e.,
$y_i=y_f=L_y/2$, in the limit $(x_i-x_f)\gg l$, the kernel obeys the power-law
dependence $K_1 \sim 1/|x_i-x_f|$. This behavior is consistent with the
power-law decay obtained from the edge-state picture for integer filling
$\nu=1$. In the interesting case of the width becoming comparable to the
magnetic length, the kernel shows oscillations with a wavelength of about $4\pi
l^2 / L_y$ which become pronounced when the two points lie on opposite edges.
This oscillatory behavior suggests that finite-size effects in realistic
situations could cause significant deviations from predictions for extended
systems.

As an application of the single-particle correlator relevant to several
experiments and proposals, we now consider tunneling between integer quantum
Hall edges caused by translational symmetry breaking along the $x$ direction.
As a simple model, we introduce localized impurities of the form \beq U (x,y)
~=~ \sum_n ~U_n ~\de (x-x_n) ~\de (y-y_n), \eeq which act as scatterers. In the
absence of a confining potential, all $k$ states are degenerate and the
scatterers cause mixing between all states. In reality, as shown in Fig.~\ref{fig:Hallbar}. b, the confining potential breaks the degeneracy, and
electrons fill states up to a Fermi momentum $k_F$ and an associated width
$L_y=2y_F$. The confining potential produces an effective electric field along
the edge, ${\cal E} = -(dV/dy)_{y=y_F}$. Electrons experience a drift velocity
given by $v_F = c |{\cal E}/B|$ and they move in opposite directions along the
top and bottom edges. Treating the scatterers within the first-order Born
approximation and assuming a linearized potential close to each edge (and thus
a linearized dispersion about the Fermi energy), we find that the scatterers
couple each $k$ state to corresponding $\pm k$ states \cite{inprep}. The
associated reflection co-efficient for a right-moving edge state $k\approx k_F$
to scatter to a left-mover $k\approx -k_F$ is given by 
\bea r &=& -
~\frac{i}{\hbar v_F}
~\sum_n ~U_n ~e^{i2k_F x_n} ~\exp \left[- ~\frac{eB}{\hbar c} ~y_n^2 \right] \non \\
& & ~\times ~\left( \frac{eB}{\pi \hbar c} \right)^{1/2} \exp \left[- ~
\frac{eB}{\hbar c} ~y_F^2 \right]. \label{rtot} 
\eea 
The reflection co-efficient is
 directly related to the matrix element for particles to tunnel between
edge states. Implicitly, it involves the single-particle propagation amplitude
to traverse from one edge to another. Our method is simple enough that it can
go beyond the strip geometry to any smooth confining potential and
configuration of tunneling sites.

The form of Eq. (\ref{rtot}) has several noteworthy features. As expected, the
tunneling matrix element for each impurity decays exponentially over a magnetic
length. For an impurity localized on an edge at a point $x_i$, tunneling to the
other edge occurs along the shortest path. The treatment here was for fermions
of charge $e$. In principle, we expect an identical form for any particle
having charge $e^*$ with this charge replacing $e$, in which case the decay of
the {\it bare} tunneling matrix element is enhanced/suppressed by a factor
$e^*/e$ in the exponent. This reasoning is consistent with derivations of
tunneling matrix elements that explicitly use the Laughlin wavefunction
\cite{Assa}. For the situation of more than one impurity, the reflection
coefficient is sensitive to interference effects coming from multiple paths. In
the case of two impurities of equal strength lying on either edge at points
$x_1$ and $x_2$, reflection processes off the two impurities have a phase
difference of $2k_F (x_2 - x_1) = 2eBy_F (x_2 - x_1)/(\hbar c)$. Thus, as
phenomenologically described in Ref. \cite{Chamon}, we explicitly see
Aharonov-Bohm interference coming from the particle traversing two different
paths enclosing a rectangular area of length $x_1-x_2$ and width $2y_F$.

In conclusion, we have derived and analyzed the form of two ubiquitous entities
- the single-particle and two-particle anyon kernels - in the physically
motivated situation of charged particles in a magnetic field in the LLL. We
have shown that the two-particle kernel in the quantum Hall bulk contains
information on statistics which is strikingly manifest in the zeros of the
kernel. We have shown that the single-particle kernel in a finite geometry
provides a faithful means of understanding features of bulk mediated tunneling
between edge state quasiparticles, such as the tunneling amplitude and
Aharonov-Bohm physics in a system with two tunneling centers. In principle,
some of our predictions for the two-particle kernel ought to translate to
realistic gate-defined Hall geometries. At the very least, our studies show
that a complete explanation of experiments that measure two-particle
properties, whether of bulk or edge-state quasiparticles will need to take into
account correlations and exclusion effects in the bulk. More spectacularly, our
studies indicate that in the future, it may be possible to perform experiments
in quantum Hall geometries, perhaps involving multi-edge tunneling, wherein
correlations show signatures of fractional statistics in angular dependences
such as those observed for fermions and bosons in scattering experiments.

We would like to acknowledge Assa Auerbach, Eduardo Fradkin and  Gordon Baym for illuminating discussions. 
This work was supported by the UIUC Department of Physics and by the  NSF under grant 
DMR 06-03528. Work in India was supported by DST  under the project SP/S2/M-11/2000. 



\begin{thebibliography}{99}
\bibitem{footnote} In this paper, our interest is restricted to Abelian fractional
statistics and Laughlin states.
\end{thebibliography}
\end{document}